\documentclass[twocolumn]{raa_twocolumn}
\usepackage{subcaption}
\usepackage{threeparttable}
\usepackage{graphicx,times}
\usepackage{natbib}
\usepackage{amssymb,amsmath}
\usepackage{enumitem}
\bibpunct{(}{)}{;}{a}{}{,}
\usepackage[pagebackref=true]{hyperref}

\begin{document}

   \title{Mock Observations for the CSST Mission: Main Surveys -- the Slitless Spectroscopy Simulation}

 \volnopage{ {\bf 20XX} Vol.\ {\bf X} No. {\bf XX}, 000--000}
   \setcounter{page}{1}
    \author{Xin Zhang \inst{1}
    \and Yue-dong Fang \inst{3}
    \and Cheng-liang Wei \inst{4}
    \and Guo-liang Li \inst{4}
    \and Feng-shan Liu \inst{1,2}
    \and Hang-xin Ji \inst{5}
    \and Hao Tian \inst{1}
    \and Nan Li \inst{1,2}
    \and Xian-min Meng \inst{1}
    \and Jian-jun Chen \inst{1}
    \and Xia Wang \inst{1}
    \and Rui Wang \inst{1}
    \and Chao Liu \inst{1,2}
    \and Zhong-wen Hu \inst{2,5}
    \and Ran Li \inst{6}
    \and Peng Wei \inst{1}
    \and Jing Tang \inst{1}
    }

    \institute{ Key Laboratory of Space Astronomy and Technology, National Astronomical Observatories, Chinese Academy of Sciences, 20A Datun Road, Beijing 100101,People’s Republic of China; {\it zhangx@bao.ac.cn}
   \and University of Chinese Academy of Sciences, Beijing, China
   \and Universitäts-Sternwarte München, Fakultät für Physik, Ludwig-Maximilians-Universität
München, Scheinerstrasse 1, 81679 München, Germany
   \and Purple Mountain Observatory, Chinese Academy of Sciences, Nanjing 210023, China
   \and Nanjing Institute of Astronomical Optics \& Technology, Chinese Academy of Sciences, Nanjing 210042, China
   \and Beijing Normal University, Beijing 100875, China
   }


\abstract{
The China Space Station Telescope (CSST), slated to become China’s largest space-based optical telescope in the coming decade, is designed to conduct wide-field sky surveys with high spatial resolution. Among its key observational modes, slitless spectral observation allows simultaneous imaging and spectral data acquisition over a wide field of view, offering significant advantages for astrophysical studies. Currently, the CSST is in the development phase and lacks real observational data. As a result, the development of its data processing pipeline and scientific pre-research must rely on the mock data generated through simulations. This work focuses on developing a simulation framework for the CSST slitless spectral imaging system, analyzing its spectral dispersing properties and structural design. Additionally, the detection performance of the slitless spectral system is assessed for various astrophysical targets. Simulation results demonstrate that nearly all 1st order spectra are accompanied by corresponding 0th order images, facilitating accurate source identification. Furthermore, the GI spectral band exhibits superior detection efficiency compared to the GV and GU bands, establishing it as the primary observational band for stellar and galactic studies. This work successfully develops a simulation framework for the CSST slitless spectroscopic equipment.
}

   \authorrunning{}            
   \titlerunning{}  
   \maketitle

%
\section{Introduction}           
\label{sect:intro}
Slitless spectroscopy is a highly efficient method for spectral detection in survey missions, offering unique advantages over traditional slit spectroscopy. Unlike slit-based observations, slitless spectroscopy removes the spatial constraints imposed by a slit, allowing all optical targets within the field of view to be dispersed onto a single detector using optical elements such as gratings, grisms, or prisms. This approach allows simultaneous spectral acquisition for all detected targets without prior source selection, significantly improving survey efficiency.

However, due to this observation method and purpose, slitless spectroscopic observations are not well-suited for high-resolution spectral observations. High-resolution spectra require more area on the detector, and the absence of slits increases the likelihood of spectral overlap, making it more challenging to achieve high resolution without significant contamination. Despite its efficiency, the absence of a slit introduces challenges, particularly spectral trace overlap in two-dimensional spectroscopic images. This overlap complicates the extraction and interpretation of individual spectra, making the detection and analysis of slitless spectra inherently more complex than direct imaging observations.

Slitless spectroscopic surveys were initially implemented in ground-based astronomical observation projects, such as the Hamburg/ESO Quasar Survey \citep{1996A&AS..115..227W} and the Quasars near Quasars Survey \citep{2008A&A...487..539W}. However, ground-based telescopes face limitations due to atmospheric interference, which affects image quality, and stronger stray light from sources such as the Moon and other terrestrial factors. In contrast, space-based observations benefit from the absence of an atmosphere and significantly reduced stray light, enabling higher-quality data acquisition. As a result, large astronomical telescopes have increasingly been deployed in space. Following the success of the Hubble Space Telescope (HST), slitless spectroscopic observations have transitioned into space-based platforms. The HST is equipped with multiple slitless spectrometers designed to address diverse scientific objectives. For example, the NICMOS/Hubble Space Telescope Grism Parallel Survey \citep{1999ApJ...520..548M} aimed to identify high-redshift galaxies by detecting their emission lines using the G141 grism ($\lambda_c$=1.5 $\mu m$, $\Delta\lambda$=0.8 $\mu m$).The GRAPES \citep{2004ApJS..154..501P} focuses on investigating high-redshift galaxies (4 < z < 7) within the Hubble Ultra Deep Field (UDF) using the grism spectrometer of the Advanced Camera for Surveys (ACS). Similarly, the PEARS survey \citep{2009AJ....138.1022S,2013ApJ...772...48P} employs the ACS G800L grism spectrometer to search for emission-line galaxies in the GOODS-N and GOODS-S fields by detecting key emission lines such as H$\alpha$, [O III], and [O II]. 3D-HST \citep{2012ApJS..200...13B} is a slitless spectroscopic survey designed to collect galaxy spectra across three-quarters of the CANDELS fields \citep{2011ApJS..197...35G, 2011ApJS..197...36K} using the WFC3/G141 and ACS/G800L grisms. This project aims to refine measurements of the cosmos and uncover evidence of galaxy formation and evolution. 

Furthermore, all major ongoing space-based optical telescope missions plan to incorporate slitless spectrometers to advance astronomical research. The Euclid \citep{2011arXiv1110.3193L} is a mission featuring a 1.2-meter space-based Korsch telescope equipped with optical and infrared wide-field imagers and slitless spectroscopy. It aims to measure the shapes and redshifts of galaxies and galaxy clusters up to redshifts of $\sim$2. The Wide Field Infrared Survey Telescope (WFIRST) \citep{2012arXiv1208.4012G}, now known as the Nancy Grace Roman Space Telescope, is the most sensitive infrared space-based telescope, featuring a 2.4-meter aperture for imaging and slitless spectroscopic observations. One of its primary missions, the Galaxy Redshift Survey (GRS) \citep{2015arXiv150303757S}, utilizes a grism to measure the angular diameter distance, DA(z), and the expansion rate,
H(z), by studying H$\alpha$ emission line galaxies at 1 < z < 2  and [O III] emission line galaxies at 2<z<3. The James Webb Space Telescope (JWST), previously known as the "Next Generation Space Telescope" (NGST) \cite{2006SSRv..123..485G}, is equipped with state-of-the-art instruments, including the Near-Infrared Imager and Slitless Spectrograph (NIRISS) \citep{2020AAS...23537211R, 2012SPIE.8442E..2RD}. NIRISS enables single-object, cross-dispersed slitless spectroscopy, providing simultaneous wavelength coverage from 0.7 to 2.5 $\mu$m.

CSST \citep{Zhan2011,2018cosp...42E3821Z,zhan2021} is a major initiative featuring a 2-meter aperture off-axis optical telescope system, scheduled to begin operations around 2027. The project offers wavelength coverage ranging from the near-ultraviolet to near-infrared bands (250 nm to 1000 nm). In addition to imaging capabilities, CSST is equipped with three bands to enable slitless spectral sky surveys. The slitless spectroscopic survey and imaging survey are designed to cover the same $\sim 40\%$ of the sky, with overlapping fields. The telescope’s field of view spans approximately 1.1 square degrees, with the slitless spectroscopic observation area accounting for 40$\%$ of this coverage. A single observation with the CSST can capture a large number of celestial objects. However, the design of the CSST spectral imaging system is highly complex, necessitating the simulation data for its slitless spectra. This simulation data is critical for supporting the development and iterative refinement of the slitless spectral data processing pipeline. Additionally, analyzing the simulation data allows for a comprehensive evaluation of the scientific capabilities and effectiveness of the CSST slitless spectral sky surveys.

This paper mainly introduces the CSST slitless spectral simulation system. Section \ref{sect:csst_sls_intro} provides an overview of CSST slitless spectrum. Section \ref{sect:main_sim_method} presents the methods of slitless spectral simulation. Section \ref{sect:result} shows the simulation results and related analyses of slitless spectra. Section \ref{sect:summary} discusses the slitless spectral simulation in future.

\section{the overview of CSST Slitless Spectroscopic Instrument}
\label{sect:csst_sls_intro}

\begin{figure}
    \centering
    \includegraphics[width=0.95\linewidth]{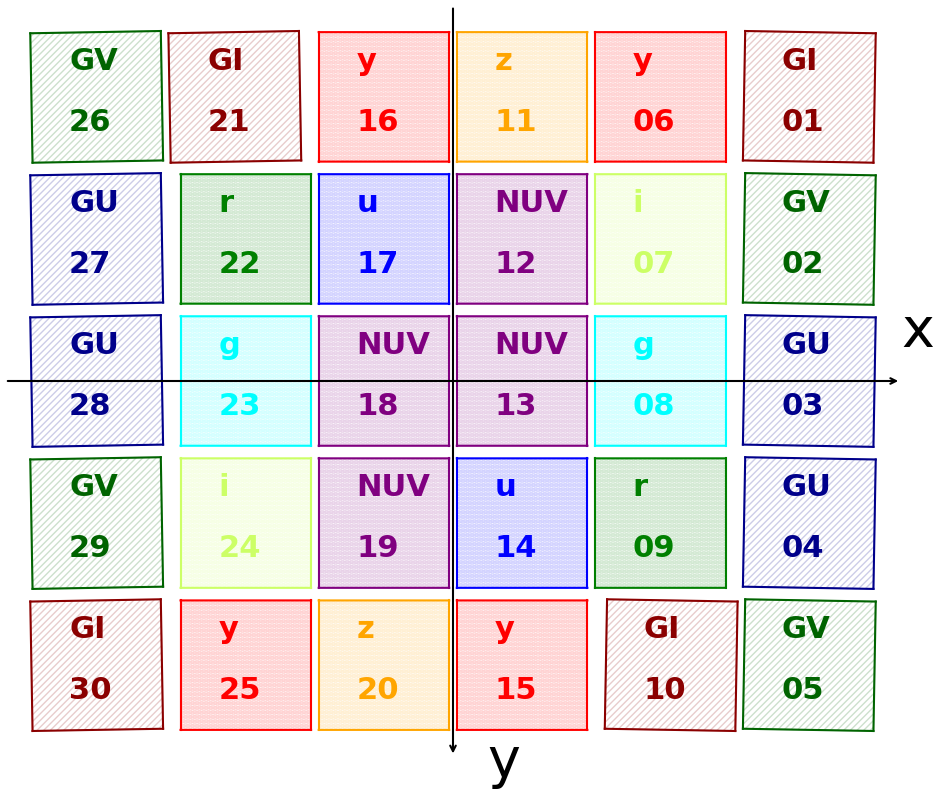}
    \caption{CSST focal plane layout. The L-shaped regions on both sides serve as the spectroscopic imaging areas, incorporating a total of 12 detectors, whereas the central region is allocated for integral imaging, comprising 18 detectors.}
    \label{fig_focal_plan}
\end{figure}

The CSST survey initiative is designed with two primary observational modes: multi-color imaging and slitless spectroscopy. These modes operate concurrently, with multi-color filters dedicated to imaging observations and diffraction gratings for slitless spectroscopy, both strategically positioned at distinct locations across the focal plane. Fig.~\ref{fig_focal_plan} illustrates the schematic layout of the CSST focal plane, which consists of 30 detector imaging areas. Of these, 18 central detectors are allocated for the multi-color imaging survey, while the remaining 12 detectors, arranged in L-shape configuration on either side, are designated for spectroscopic imaging.

\begin{figure*}[htbp]
    \centering
    \begin{subfigure}[b]{0.45\textwidth}
        \includegraphics[width=\textwidth]{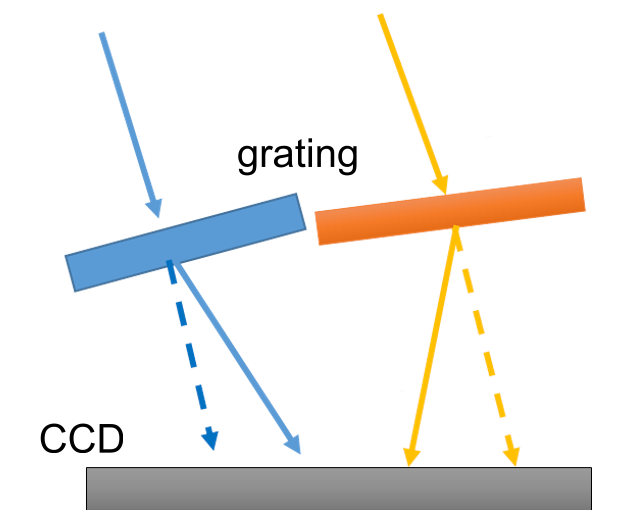}
        \caption{}
        \label{fig_double_grating1}
    \end{subfigure}
    \hspace{5pt}
    \begin{subfigure}[b]{0.45\textwidth}
        \includegraphics[width=\textwidth]{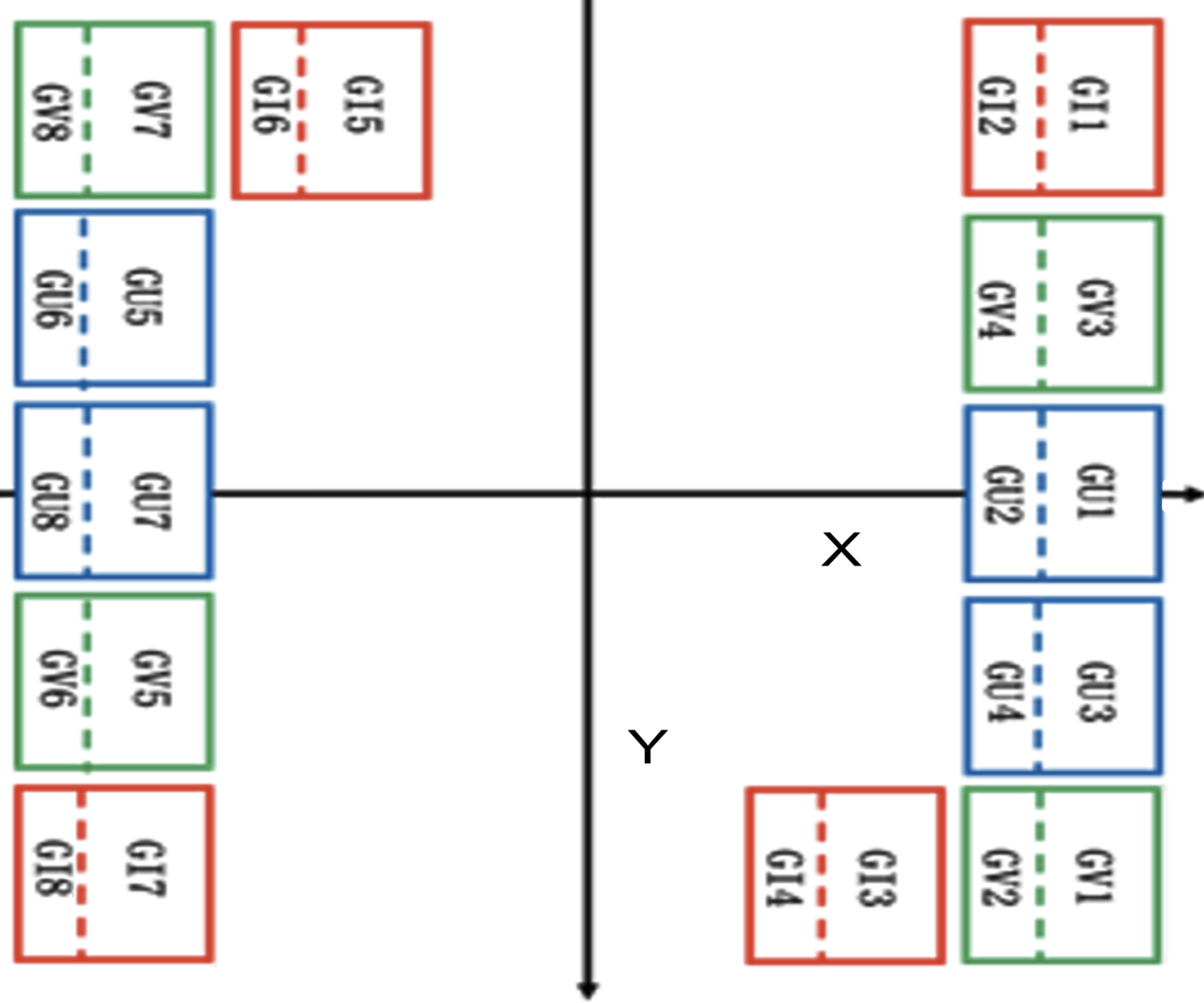}
        \caption{}
        \label{fig_double_grating2}
    \end{subfigure}
    \caption{CSST grating structure.The two gratings disperse towards the center avoid the lost of 1st order spectrum. (a) illustrates a schematic diagram of a slitless spectroscopic imaging region, which consists of a single detector and two opposing diffraction gratings designed for spectral dispersion. Solid lines indicate the propagation direction of light altered by the grating, while dashed lines indicate the propagation direction of light without the grating.(b) depicts the layout of the slitless spectroscopic imaging region on the focal plane. Each detector is divided into two segments due to the presence of the two opposing diffraction gratings positioned above it.}
    \label{fig_double_grating}
\end{figure*}

The CSST slitless spectrograph uses diffraction gratings as the dispersive elements because of their simple design and high resolution capability. Tab.~\ref{Tab_design_param} shows CSST slitless spectrograph design parameters.In the design, the primary working order of the slitless spectrograph is the 1st order spectrum. The gratings for the slitless spectrograph are positioned approximately 70 millimeters  above the detectors. Based on the characteristics of the gratings and their dispersive properties, the entire grating disperses the 1st order spectrum in a single direction. This results in some of the 1st order spectrum being dispersed outside the detector area. To ensure that the 1st order spectrum is dispersed within the detector as much as possible, the CSST slitless spectrograph grating structure features an innovative design. Each grating assembly above the detectors consists of two gratings, with each grating dispersing towards the center of the detector. This design guarantees that all 1st order spectra are imaged within the detector. Fig.~\ref{fig_double_grating} shows this grating structure.

\begin{figure}
    \centering
    \includegraphics[width=0.95\linewidth]{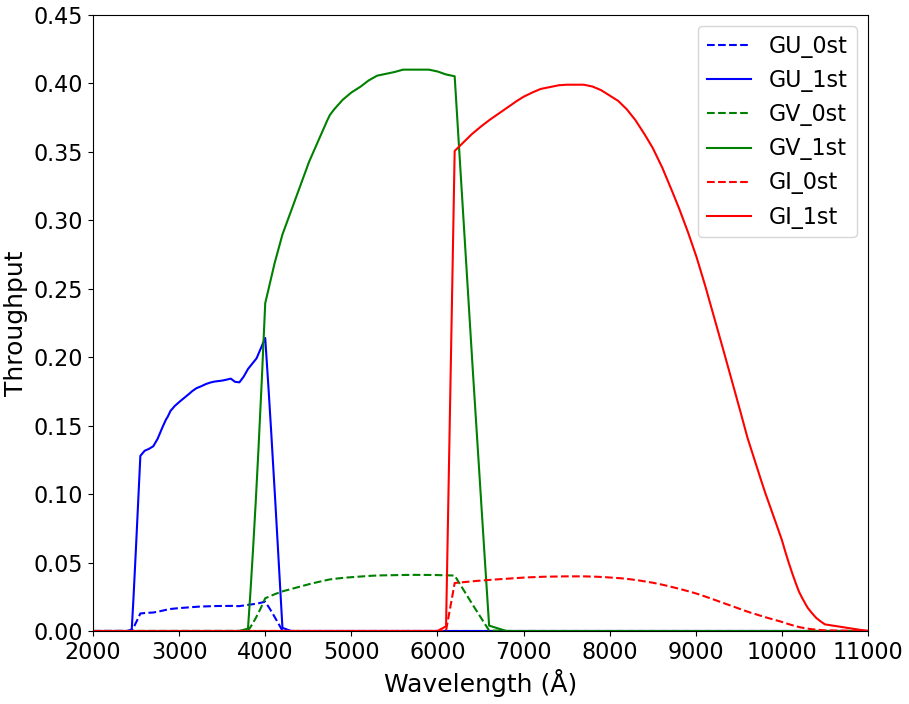}
    \caption{The throughput of 0st order and 1st order spectrum of CSST three bands, including optical transmission, filter throughput, grating diffraction efficiency and the quantum efficiency (QE) of the CCD. The throughput of the  0st order spectrum is globally scaled according to the efficiency of the 1st order spectrum, adjusted to 10\% of the 1st order spectral efficiency.}
    \label{fig_throughput}
\end{figure}

The primary orders of the slitless spectrograph are the 0th order image and the 1st order spectrum. For the 1st-order spectrum, the minimum average efficiency requirements are set at $\geq$ 60\% for the GU band and $\geq$ 65\% for the GV and GI bands. The efficiency of the 0th order image is 5\% to 10\% of the efficiency of the 1st order spectrum. Note that the above parameters are design specifications; the actual measured efficiency will be subsequently updated in the software of CSST Main Survey Simulator \footnote{\url{https://csst-tb.bao.ac.cn/code/csst-sims/csst_msc_sim}}. Fig.~\ref{fig_throughput} illustrates the throughput of the CSST gratings for the 0th order and 1st order spectra, including optical transmission, filter throughput, grating diffraction efficiency and the quantum efficiency (QE) of the CCD.

\begin{table}[htbp]
\begin{threeparttable}
\centering
\caption{The design parameter of CSST slitless spectrograph}
\label{Tab_design_param}
\begin{tabular}{|c|c|c|c|c|} 
\hline
\textbf{Band} & \textbf{Central(nm)}\tnote{a} & \textbf{Min(nm)}\tnote{b} & \textbf{Max(nm)}\tnote{b} & \textbf{R}\tnote{c} \\ \hline
GU & 337 & 255 & 410 & 287 \\ \hline
GV & 525 & 400 & 640 & 232 \\ \hline
GI & 810 & 620 & 1000 & 207 \\ \hline
\end{tabular}
\begin{tablenotes}
\item[a] Central wavelength of the spectral band.
\item[b] Designed maximum and minimum wavelengths within each spectral band.
\item[c] R denotes the spectral resolution at the central wavelength.
\end{tablenotes}
\end{threeparttable}
\end{table}

\section{Method for Slitless Spectroscopic Simulation Based on Optic System}
\label{sect:main_sim_method}

\subsection{The principle of grating spectroscopy}

\label{sec:grating_principle}

From Section~\ref{sect:csst_sls_intro} We can know that diffraction gratings with regular patterns is the optical components of the CSST slitless spectrograph. The light diffracted by a grating is found by summing the light diffracted from each of the elements, and is essentially a convolution of diffraction and interference patterns. So all gratings have intensity maxima at angles $\theta_m$ which are given by the grating equation
\begin{equation}
  \label{equ:Diffraction}
  m \lambda = d (sin \theta_i \pm sin \theta_m)
\end{equation}  
where $\theta_i$ is the angle at which the light is incident, $\theta_m$ is the diffraction angles, $d$ is the separation of grating constant, and $m$ is the number of spectrum order which is integer. The sign "$\pm$" in the equation depends on the relative positions of the incident and diffracted rays relative to the grating normal: the sign of "-" applies when both rays reside on the same side of the normal, whereas the sign of "+" is used when they lie on opposite sides.

\begin{figure}
    \centering
    \includegraphics[width=0.95\linewidth]{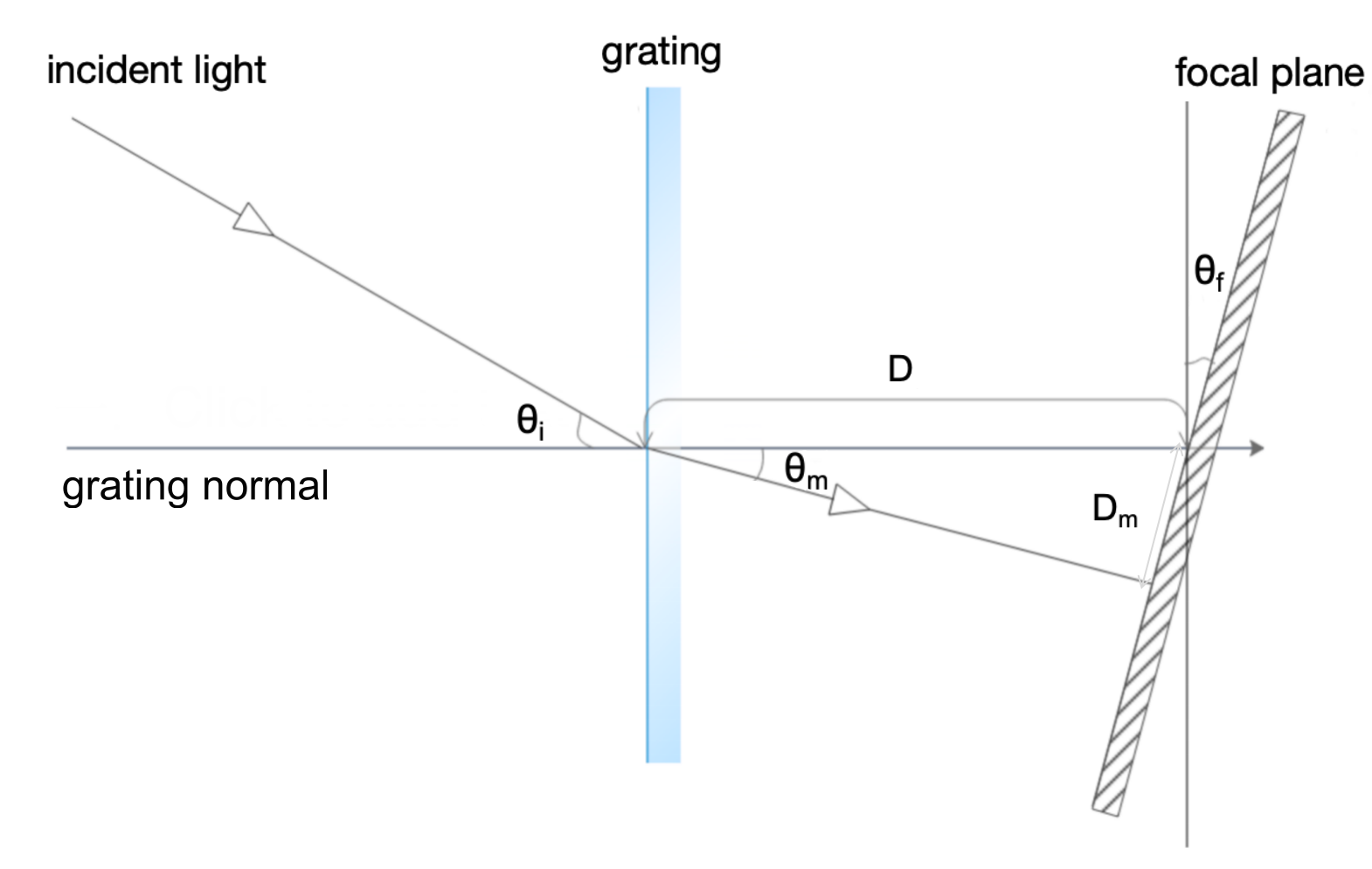}
    \caption{the Schematic Diagram of CSST Slitless Spectrograph Optical Path. $\theta_i$ is the incident angle, defined as the angle between the incident light ray and the grating normal; $\theta_m$ is the diffraction angle, defined as the angle between the diffracted light ray and the grating normal; $\theta_f$ is the grating tilt angle, defined as the angle between the grating surface and the focal plane; $D$ represents the average distance between the grating and the focal plane, $D_m$ is the distance between the diffraction point and the origin where the grating normal intersects the focal plane.}
    \label{fig_sls_pos}
\end{figure}

Based on the optical path design of the CSST, it is evident that the incident light strikes the grating at a specific angle relative to its normal, while the grating itself is oriented at a distinct angle relative to the focal plane, resulting in a non-parallel configuration between these components. Fig.~\ref{fig_sls_pos} shows the sketch of CSST slitless spectrograph optical path. In this sketh, the definition of $\theta_m$ and  $\theta_i$ is the same as described above, $D$ represents the average distance between the grating and the focal plane, $D_m$ is the distance between the diffraction point and the origin, $\theta_f$ shows the angle between the focal plane and the grating plane. According to Equation~\ref{equ:Diffraction} and the geometric relationship in Fig.~\ref{fig_sls_pos}, $D_m$ can be expressed as:
\begin{equation} 
\label{equ:optic_path}
  D_m = \frac{ D \, ( m \lambda - d \sin \theta_i ) }{ d \cos(\theta_m - \theta_{\mathrm{f}}) } 
\end{equation}
In the equation, $\theta_i$, $\theta_m$, $\theta_f$, $D$ and $D_m$ follow Equation~\ref{equ:optic_path} definitions, while $d$ denotes the grating constant. For the CSST grating design, this constant (d) is determined by the groove density. $D_m$ represent the position of the light with wavelength $\lambda$ in the detector. From Equation~\ref{equ:optic_path}, we can find that the position $D_m$ is only related to the wavelength, giving the optical path remains unchanged.Furthermore, $D_m$ is linearly related to the wavelength.In addition, if the lines of the grating are parallel and uniform, the spectrum obtained by the dispersion will also appear as a straight line. In the subsequent simulations, we will also assume that the relationship between x and y coordinates on the detector, where the spectrum is imaged, is linearly correlated based on this premise.

\subsection{The representation methods for slitless spectra}
\begin{figure}
    \centering
    \includegraphics[width=0.95\linewidth]{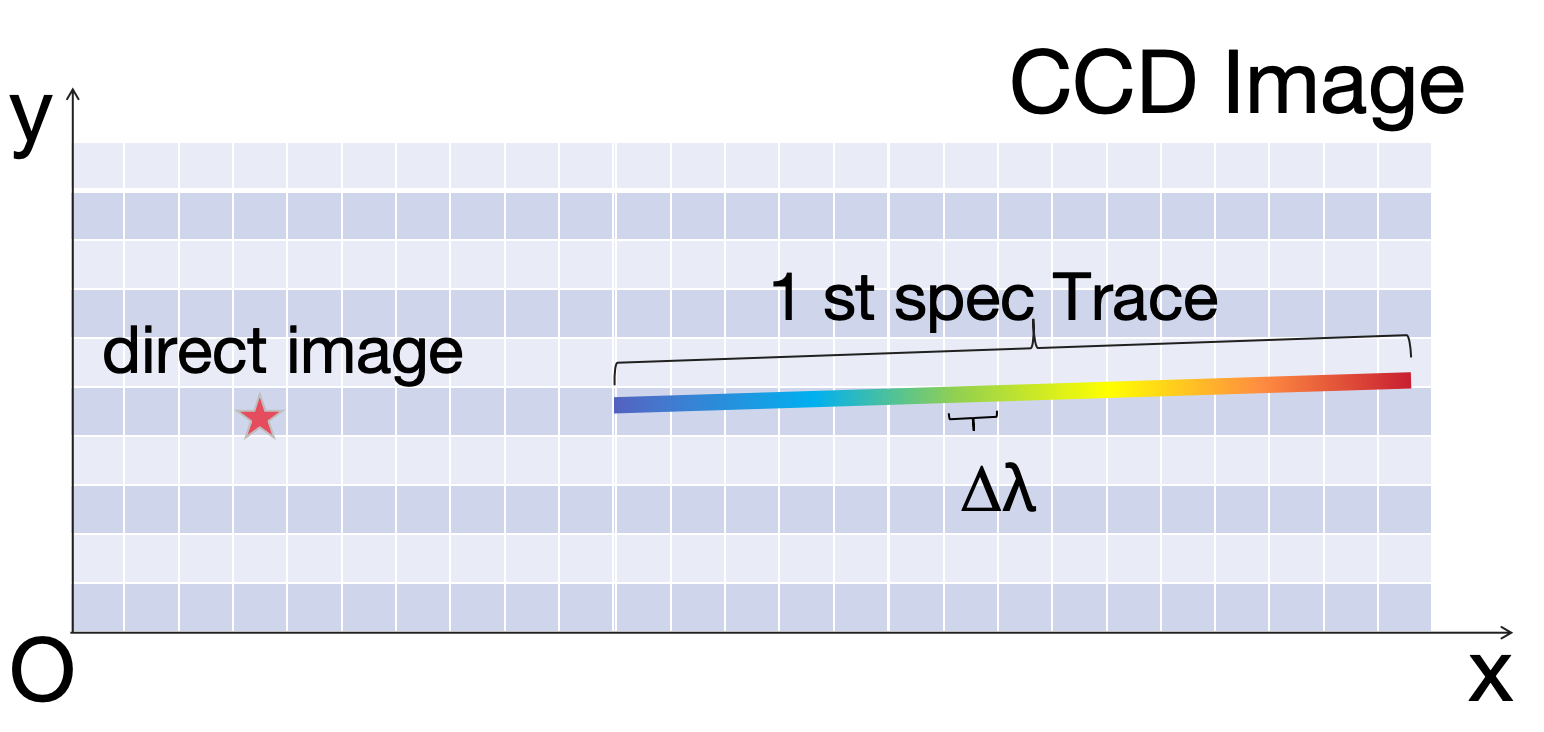}
    \caption{Schematic diagram of spectral dispersion image. This schematic diagram illustrates the pixel grid of the detector. The red star indicates the directly imaged position on the detector (virtual image), while the colored lines represent the dispersed first-order spectrum across the detector surface. Each pixel integrates photons within a wavelength interval $\Delta \lambda$.}
    \label{fig_sls_disp}
\end{figure}

From Section~\ref{sec:grating_principle}, We can conclude that the position of the spectrum image along the y-axis ($y$) in relation to the x-axis ($x$), as well as the position $D_m$ with respect to the wavelength $\lambda$, exhibit a linear relationship. Fig.~\ref{fig_sls_disp} provides a schematic illustration of the spectrum being imaged on the detector. In the diagram, the direct image serves as a virtual image, which is utilized as a reference point for wavelength calibration. We can use a polynomial to describe the positional relationship of the spectrum  as Equation~\ref{equ:pos_relation_1}. 

\begin{equation}
  \label{equ:pos_relation_1}
  y = a_0 + a_1x+a_2x^2+......+a_nx^k
\end{equation}
Based on the above description, the spectrum trace can be represented by a 1st  order polynomial, so the spectrum trace can described by 
\begin{equation}
  \label{equ:pos_relation_2}
  y = a_0 + a_1x
\end{equation}
In addition, the trajectory of the spectrum is also related to the incident position of light.Here, we believe that the trajectory of spectral dispersion is related to the position where light directly falls on the detector (that is, the position of direct imaging). So we can use the direct image position (m, n) to describe the coefficients of spectrum trajectory $a_0$ and $a_1$ as follow:
\begin{equation}
  \label{equ:pos_efficiency_a0}
  a_k = b_{k,0} + b_{k,1}m + b_{k,2}n + b_{k,3}m^2 + b_{k,4}n^2 + b_{k,5}mn  
\end{equation}
so
\begin{equation}
  \label{equ:pos_relation_comb}
  \begin{aligned}
  y = & b_{0,0} + b_{0,1}m + b_{0,2}n + b_{n0,3}m^2 + b_{0,4}n^2 + b_{0,5}mn + \\
  &(b_{1,0} + b_{1,1}m + b_{1,2}n + b_{1,3}m^2 + b_{1,4}n^2 + b_{1,5}mn)x
  \end{aligned}
\end{equation}
From experience, the coefficients of the spectral trajectory are well-represented by the aforementioned two-dimensional second-order polynomial.

Similarly, we can use a polynomial to describe the relationship between the wavelength of the spectrum and the spectral position.

\begin{equation}
  \label{equ:wave_pos_relation_1}
  \lambda = \alpha_0 + \alpha_1 L_{trace}+\alpha_2 L_{trace}^2+......+\alpha_n L_{trace}^k
\end{equation}
$L_{trace}$ represents the length of spectral trajectory in the spectral position (x, y), can be described by 

\begin{equation}
  \label{equ:wave_pos_dL}
  L_{trace} = \int_0^X {\sqrt{dx^2+dy^2}}, y = f(x)
\end{equation}
Similarly, we can use a polynomial to describe the relationship between the wavelength of the spectrum and the spectral position. So the Equation \ref{equ:wave_pos_relation_1} can be removed the higher-order terms as 

\begin{equation}
  \label{equ:wave_pos_relation_2}
  \lambda = \alpha_0 + \alpha_1 L_{trace}
\end{equation}
The coefficients $\alpha_0$ and $\alpha_0$ can be described by a position-dependent polynomial like Equation \ref{equ:pos_efficiency_a0} as

\begin{equation}
  \label{equ:wave_efficiency_a0}
  \alpha_k = \beta_{k,0} + \beta_{k,1}m + \beta_{k,2}n + \beta_{k,3}m^2 + \beta_{k,4}n^2 + \beta_{k,5}mn  
\end{equation}
By combining Equation \ref{equ:wave_efficiency_a0} and Equation \ref{equ:wave_pos_relation_1}, we can represent lambda with the following equation
\begin{equation}
  \label{equ:wave_pos_relation_comb}
  \begin{aligned}
  & \lambda  =  \beta_{0,0} + \beta_{0,1}m + \beta_{0,2}n + \beta_{0,3}m^2 + \beta_{0,4}n^2 + \beta_{0,5}mn+\\
  &(\beta_{1,0} + \beta_{1,1}m + \beta_{1,2}n + \beta_{1,3}m^2 + \beta_{1,4}n^2 + \beta_{1,5}mn) L_{trace}
  \end{aligned}
\end{equation}

\subsection{Fitting the trajectory of spectra}
\begin{figure}
    \centering
    \includegraphics[width=0.95\linewidth]{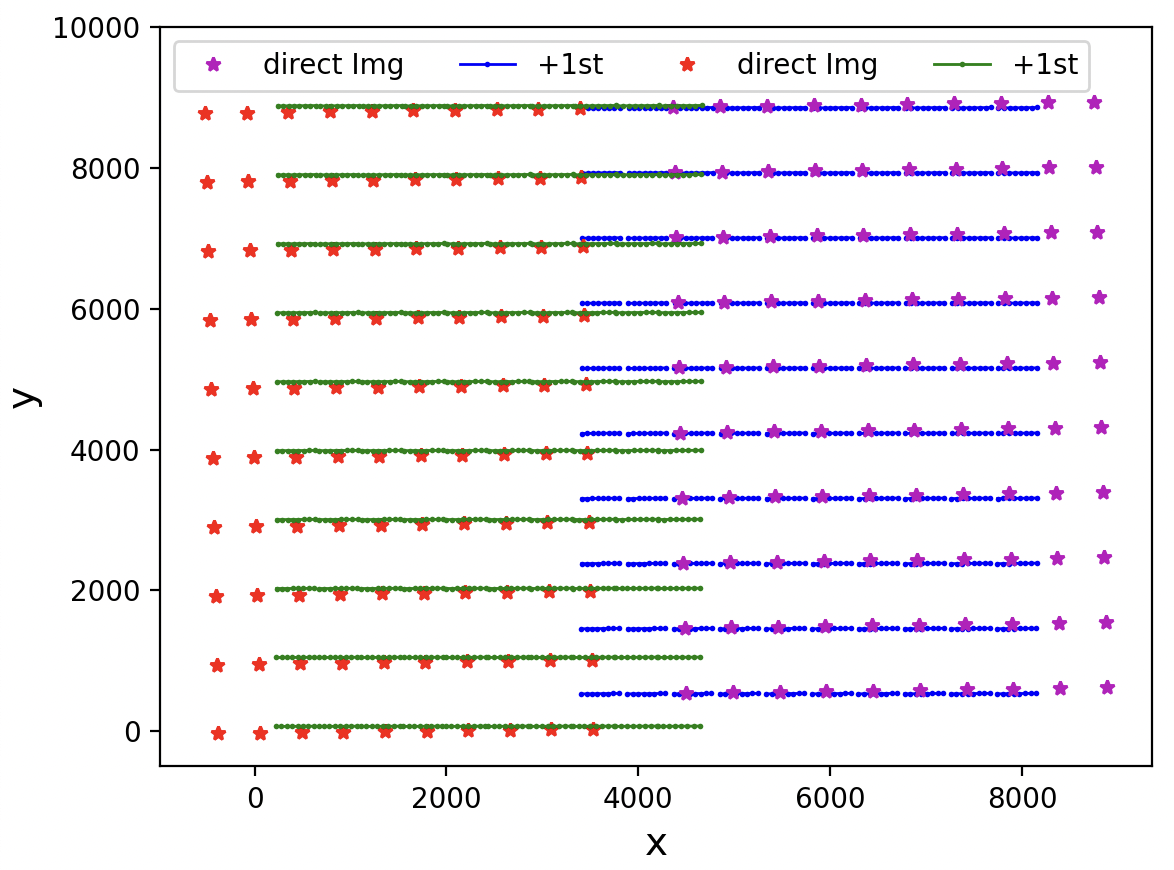}
    \caption{the Points of Optical Simulation.For each grating, 10$\times$10 positions are selected. The stars in the figure are the positions of direct imaging. The dotted lines are the 1st order spectra. Each direct imaging point corresponds to a 1st order spectrum. Each 1st order spectrum is uniformly composed of eight points. }
    \label{fig_optical_sim_pos}
\end{figure}

As shown in the previous section, the trajectory of the spectrum can be described by Equation~\ref{equ:pos_relation_comb}, ~\ref{equ:wave_pos_relation_comb} which are both related to the imaging position (m, n). So we need to obtain the spectral trajectory characteristics of imaging at different positions on the detector. Since there is no measured observation data, we take the data from optical simulation as the input for fitting the spectral trajectory. The optical simulation includes the black box model of the main optical path of the optical system (Zhang Ban, and Xiaobo Li, et al., in preparation) and the design of the CSST slitless spectrometer grating. Here, it is necessary to simulate the changes in imaging spectra at different positions on the entire detector for the grating. Therefore, we have designed relatively dense sampling. And due to the unique design of the CSST slitless spectrometer, as we can see from Fig.~\ref{fig_double_grating}, two gratings cover one detector. Therefore, we have selected 10$\times$10 sampling points within the range of a single grating. Through optical simulation, we obtain the spectral trajectory characteristics of 100 positions. For each spectral trajectory, 8 wavelength are uniformly selected for representation. The working orders of the CSST slitless spectrum are the 0th-order image and the 1st order spectrum. In addition, we have added three non-working order spectra in the simulation - the 2nd, -1st, and +2nd orders. Therefore, the optical simulation performs a total of five order spectral simulations (0th, $\pm$1st, and $\pm$2nd) for each position. Fig.~\ref{fig_optical_sim_pos} shows the positions of the optical simulation sampling points. This figure takes GV as an example. GU and GI are similar to this. The figure only shows the positions of the 1st order spectra. The other orders are similar to this, except that the dispersion direction and position are slightly different.

For each grating, the data of the 10$\times$10 positions as mentioned above are used as input, and the 12 parameters in Equation \ref{equ:pos_relation_comb} and Equation \ref{equ:wave_pos_relation_comb} need to be fitted respectively. We use the least squares method to fit the 12 parameters \( \Vec{b} \) -
\[
\left[ b_{0,0}, b_{0,1}, b_{0,2}, b_{0,3}, b_{0,4}, b_{0,5}, b_{1,0}, b_{1,1}, b_{1,2}, b_{1,3}, b_{1,4}, b_{1,5} \right]
\]
representing the spectral position and the 12 parameters \( \Vec{\beta} \) -
\[
\left[ \beta_{0,0}, \beta_{0,1}, \beta_{0,2}, \beta_{0,3}, \beta_{0,4}, \beta_{0,5}, \beta_{1,0}, \beta_{1,1}, \beta_{1,2}, \beta_{1,3}, \beta_{1,4}, \beta_{1,5} \right]
\]
 representing the relationship between spectral wavelength and position. For the fitting of spectral position parameters, Equation \ref{equ:fitting_pos_least_squares} is constructed.
\begin{figure*}
\begin{equation}
  \label{equ:fitting_pos_least_squares}
\overbrace{
\left[
\renewcommand{\arraystretch}{0.8}
\begin{array}{c}
y_0\\ y_1\\ .\\ .\\ .\\ .\\ .\\ .\\ .\\ .\\ y_{i-1}\\ y_i
\end{array}
\right]
}^{Y}
=
\overbrace{
\left[
\renewcommand{\arraystretch}{0.8}
\begin{array}{cccccccccccc}
1 & m & n & m^2 & n^2 & mn & x_0 & mx_0 & nx_0 & m^2x_0 & n^2x_0 & mnx_0 \\
1 & m & n & m^2 & n^2 & mn & x_1 & mx_1 & nx_1 & m^2x_1 & n^2x_1 & mnx_1 \\
&&&&&&.&.\\
&&&&&&.&.\\
&&&&&&.&.\\
&&&&&&.&.\\
&&&&&&.&.\\
&&&&&&.&.\\
&&&&&&.&.\\
&&&&&&.&.\\
1 & m & n & m^2 & n^2 & mn & x_{i-1} & mx_{i-1} & nx_{i-1} & m^2x_{i-1} & n^2x_{i-1} & mnx_{i-1}\\
1 & m & n & m^2 & n^2 & mn & x_i & mx_i & nx_i & m^2x_i & n^2x_i & mnx_i
\end{array}
\right]
}^{X}
\overbrace{
\left[
\renewcommand{\arraystretch}{0.8}
\begin{array}{c}
b_{0,0}\\ b_{0,1}\\ b_{0,2}\\ b_{0,3}\\ b_{0,4}\\ b_{0,5}\\ b_{1,0}\\ b_{1,1}\\ b_{1,2}\\ b_{1,3}\\ b_{1,4}\\ b_{1,5}
\end{array}
\right]
}^{b}
\end{equation}
\end{figure*}
X represents the data matrix, b is the parameter vector and Y is the data result related to X. (m, n) is the position where the light directly enters the focal plane (or detector), ($x_i, y_i$) denotes the position of spectra whose origin is the position (m,n) where the light enters the focal plane. The least squares method Equation \ref{equ:fitting_pos_least_squares_simple} is used to fit the parameter vector b. The one with the smallest residual is selected as the fitting result. The fitting method for the parameter vector $\Vec{\beta}$ is the same as that for $\Vec{b}$. Refer to the fitting process of $\Vec{b}$ for details.

\begin{equation}
  \label{equ:fitting_pos_least_squares_simple}
  \underset{\Vec{b}}{min}
  =||Xb-Y||_2
\end{equation}

\begin{figure*}
    \centering
    \includegraphics[width=0.95\linewidth]{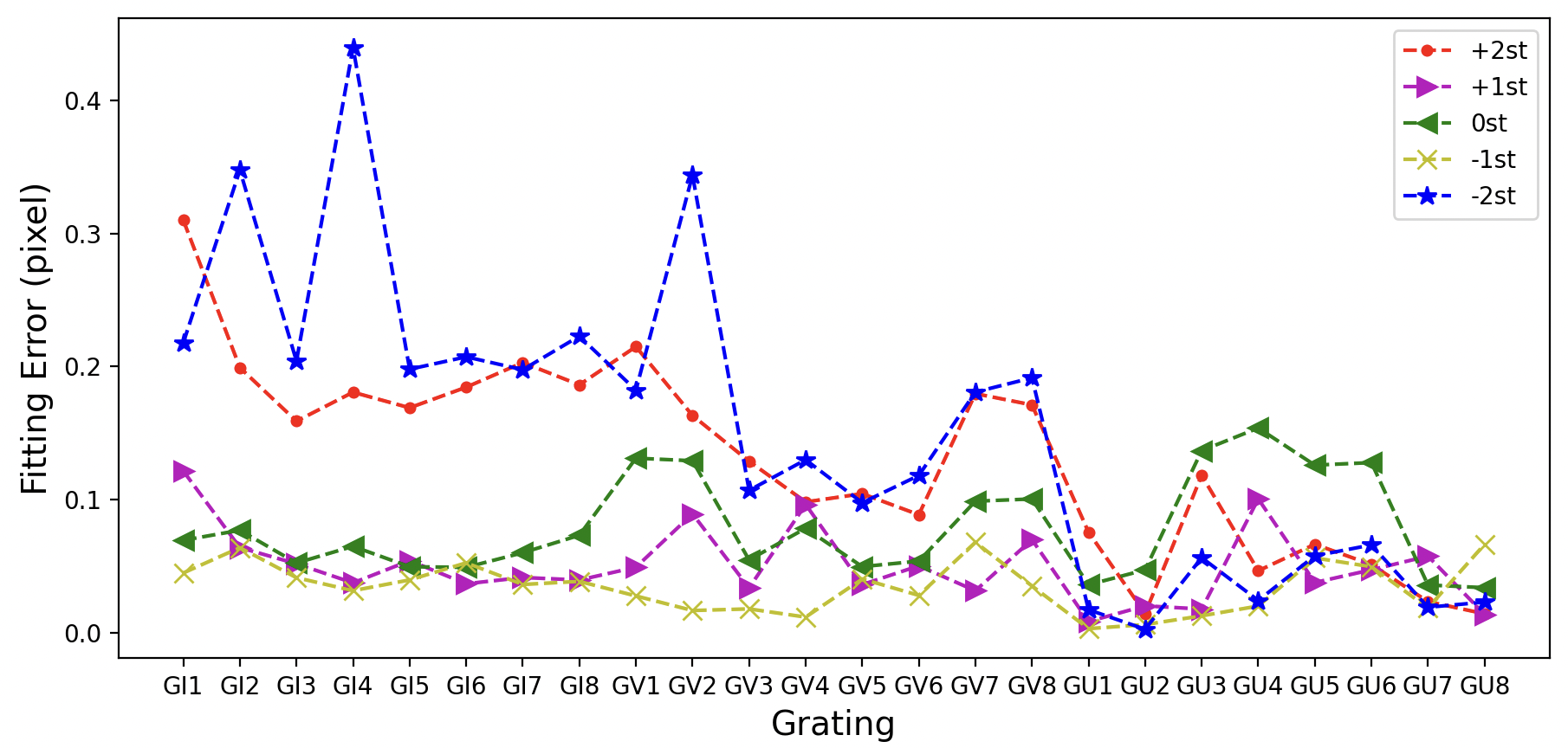}
    \caption{ The fitting result of the spectra position. The x of every spectrum of the 10$\times$10 position every grating as input and use fitting polynomial calculating the result y compare with the input y. Calculate the error of all y values on each grating to obtain the fitting error of each grating. }
    \label{fig_fitting_pos_result}
\end{figure*}

To verify the fitting results, the following procedure was adopted. For each grating, the x-coordinates at the 10$\times$10 input positions were used as inputs to the fitting polynomial, and the corresponding y-values were calculated. These calculated y-values were then compared with the true y-values at each position. The statistical error between the 100 calculated y-values and the true values was computed for each grating. As illustrated in Fig.~\ref{fig_fitting_pos_result}, the error was found to be less than 0.2 pixels for both the 0th and 1st order spectra at the working level.

\section{Simulation Result}
\label{sect:result}

\begin{figure*}[htbp]
    \centering
    \begin{subfigure}[b]{0.3\textwidth}
        \includegraphics[width=\textwidth]{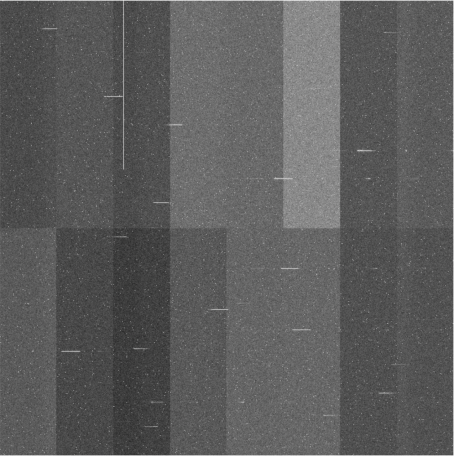}
        \caption{GU}
        \label{fig_sim_result1}
    \end{subfigure}
    \hspace{5pt}
    \begin{subfigure}[b]{0.3\textwidth}
        \includegraphics[width=\textwidth]{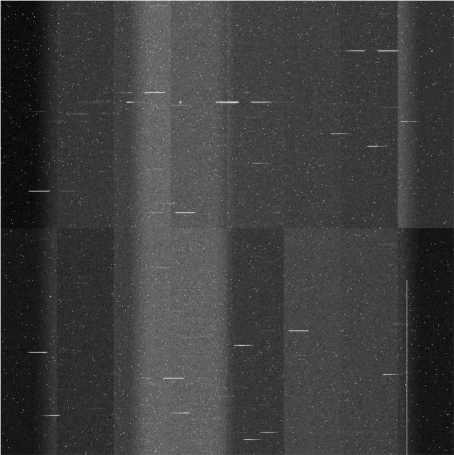}
        \caption{GV}
        \label{fig_sim_result2}
    \end{subfigure}
    \hspace{5pt}
    \begin{subfigure}[b]{0.3\textwidth}
        \includegraphics[width=\textwidth]{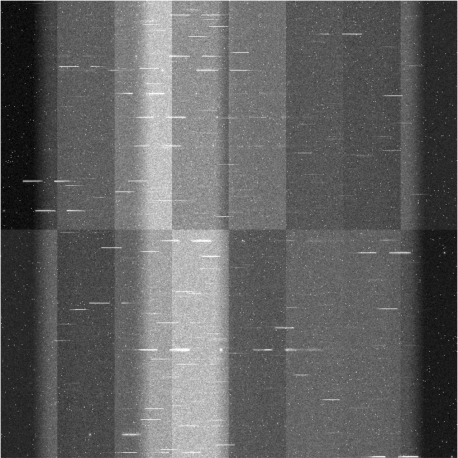}
        \caption{GI}
        \label{fig_sim_result3}
    \end{subfigure}
    \caption{CSST slitless spectrum image simulation result. Image size is 9k$\times$9k.}
    \label{fig_sim_result}
\end{figure*}

\begin{figure*}[htbp]
    \centering
    \begin{subfigure}[c]{0.45\textwidth}
        \centering
        \includegraphics[width=\textwidth]{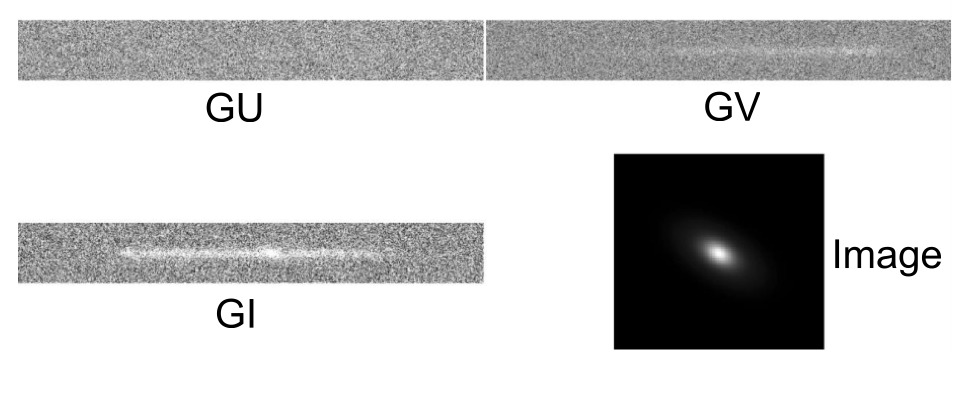}
        \caption{single source spectrum image}
        \label{fig_sim_result1}
    \end{subfigure}
    \hspace{5pt}
    \begin{subfigure}[c]{0.45\textwidth}
        \includegraphics[width=\textwidth]{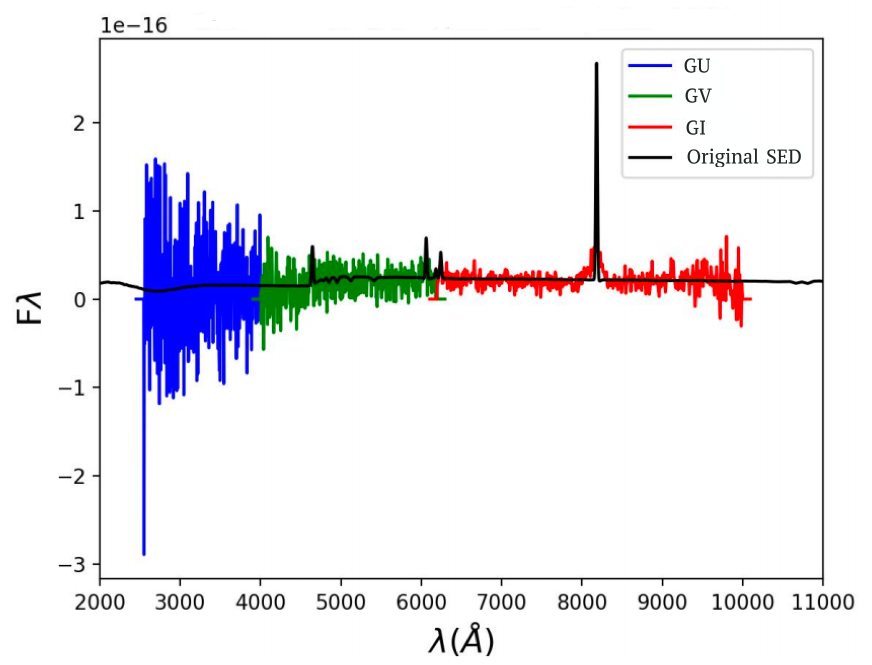}
        \caption{1d spectrum extracted}
        \label{fig_sim_result2}
    \end{subfigure}
    \caption{Presentation of spectrum extracted from simulation data. (a) includes GU, GV, GI and direct image of a galaxy. (b) shows the extracted result.}
    \label{fig_sim_result_extract}
\end{figure*}

Through the description of the above sections, we have shown the method of spectral description. For the spectral description of different orders, we have adopted the description method of the aXe extraction software\citep{2009PASP..121...59K}. Each grating corresponds to a configuration file, which contains the polynomial parameters of the spectral trajectory of each order and the light transmission efficiency of the spectrum. For detailed description, see the CSST simulation software\footnote{\url{https://csst-tb.bao.ac.cn/code/csst-sims/csst_msc_sim}}. Using the configuration files of different gratings and the efficiency files of different spectra orders, we can generate spectral images of different orders of the target source on the image. By adding the sky light, cosmic rays and CSST instrument effects (including background, dark current, gain of 16-channel readout, nonlinearity, brighter-fatter effect, PRUN, bad pixels, bad pixel columns, blooming, etc.) to this image, we can obtain spectral images under different gratings. From the image results, it can be seen that the image contains the unique bidirectional spectroscopic characteristics of CSST slitless spectra. ~\ref{fig_sim_result} presents the results of the CSST slitless spectral simulation images, displaying the simulation outcomes for detectors corresponding to GU, GV, and GI. The results reveal that the edges of the images exhibit slight darkening, a phenomenon attributed to dispersion. Specifically, at the image edges, a portion of the light is dispersed beyond the detector's boundaries. In addition, it can also be seen from the image that there will be a bright strip in the area slightly to the left of the middle of the image. This phenomenon is particularly obvious in the GI band. This is because of the design of bidirectional spectroscopy, and the sky light is dispersed to the middle. Fig.~\ref{fig_sim_result_extract} shows the cutout of the two-dimensional spectra of three bands of a galaxy. Extract these three spectra and compare them with the input spectra. As can be seen from the figure, the spectra extracted from the simulation data can be well matched with the input spectra.

In the aforementioned measurements, we rely on two ideal assumptions: first, that the reference point for the measurements (either the direct imaging position or the position of the 0th order image) is precisely known, and second, that the spectral dispersion profile used in the simulations is directly applicable. These two pieces of information are critical for spectral extraction, and by avoiding measurement errors in these aspects, we are able to achieve nearly perfect spectral extraction results. In practical data processing, the accuracy of wavelength positioning largely depends on the precision of the spectral dispersion profile construction and the centering accuracy of the reference point. Assuming that one resolution element of the slitless spectrum spans 4 pixels (with R80 = 0.3$^{\prime\prime}$ and a pixel size of 0.074$^{\prime\prime}$), and if the combined precision of these two factors is 0.5 pixels, the wavelength deviations for the GU, GV, and GI bands, calculated based on the data in Tab.~\ref{Tab_design_param}, would be 1.47 $\text{\AA}$, 2.83 $\text{\AA}$, and 4.89 $\text{\AA}$, respectively. Taking the central wavelength of the GI band (8100 $\text{\AA}$) as an example, if the rest wavelength is 5000 $\text{\AA}$ and the measurement deviation is 4.89 $\text{\AA}$, the combined errors from reference point centering and spectral dispersion profile measurement would lead to an impact on the spectroscopic redshift precision of approximately 1$‰$. This estimate does not yet include the more complex instrumental effects, underscoring that precise spectral extraction is crucial for achieving accurate photometric redshift measurements in practical observations.

Using Galaxia as the input of the stellar catalog and CosmoDC2 as the input of the galaxy catalog, a region of approximately 0.4 square degrees is selected for statistics in the central region with RA and dec of [60, -40]°. In this definition, if the signal-to-noise ratio of a single pixel in the observation area of the target source is greater than 1, it is considered that the target position is an observable source. From the above definition, we can obtain the results as shown in Tab.~\ref{Tab_number_density}, the results here are all from simulations with a single exposure of 150 seconds. As shown in the table, nearly all detectable 1st order spectra have corresponding 0th order images detectable, except in cases where the 0th order image falls outside the detector's field of view. The detection efficiency of GU is much lower than that of GV and GI. In addition, due to galaxy spectra, the detection of galaxies in the GI band is much higher than in the other two bands.

\begin{table*}[htbp]
\centering
\caption{Number density estimation of galaxies and stars under CSST slitless spectrum imaging}
\label{Tab_number_density}
\begin{tabular}{|c|c|c|c|} 
\hline
 & \textbf{GU (/$arcmin^2$)} & \textbf{GV(/$arcmin^2$)} & \textbf{GI(/$arcmin^2$)}  \\ \hline
0st Order (total) & 1.11 & 3.56 & 7.43 \\ \hline
1st Order (total) & 0.82 & 1.99 & 4.32 \\ \hline
1st Order (stellar) & 0.81 & 1.68 & 1.96  \\ \hline
1st Order (galaxy) & 0.01 & 0.31 & 2.36 \\ \hline
\end{tabular}
\end{table*}

\begin{figure*}[htbp]
    \centering
    \begin{subfigure}[b]{0.45\textwidth}
        \includegraphics[width=\textwidth]{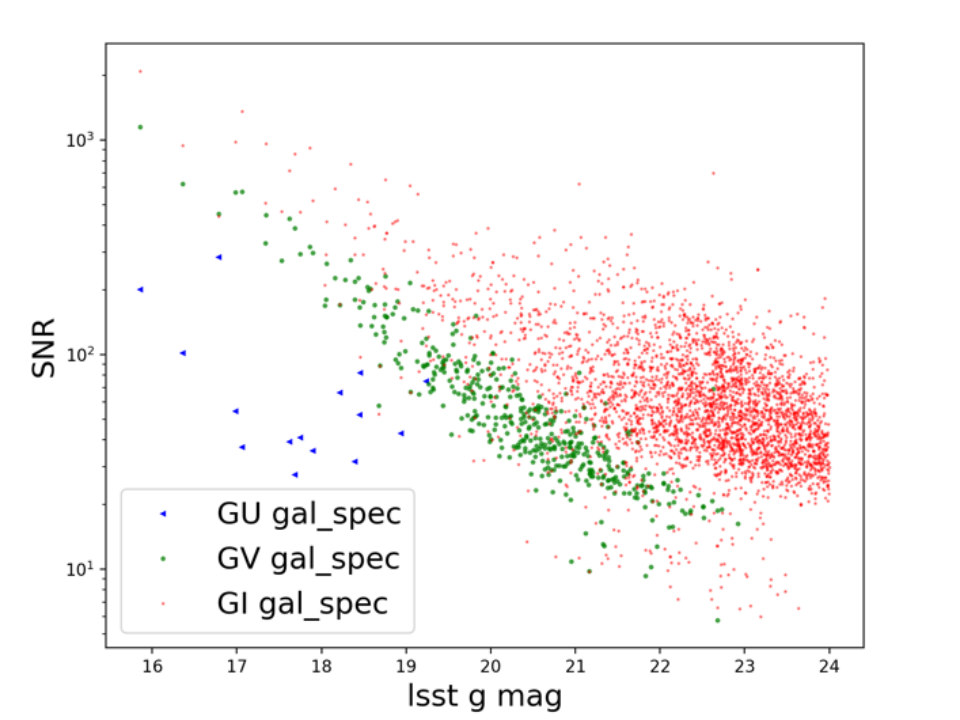}
        \caption{galaxy}
        \label{fig_sim_snr1}
    \end{subfigure}
    \hspace{5pt}
    \begin{subfigure}[b]{0.45\textwidth}
        \includegraphics[width=\textwidth]{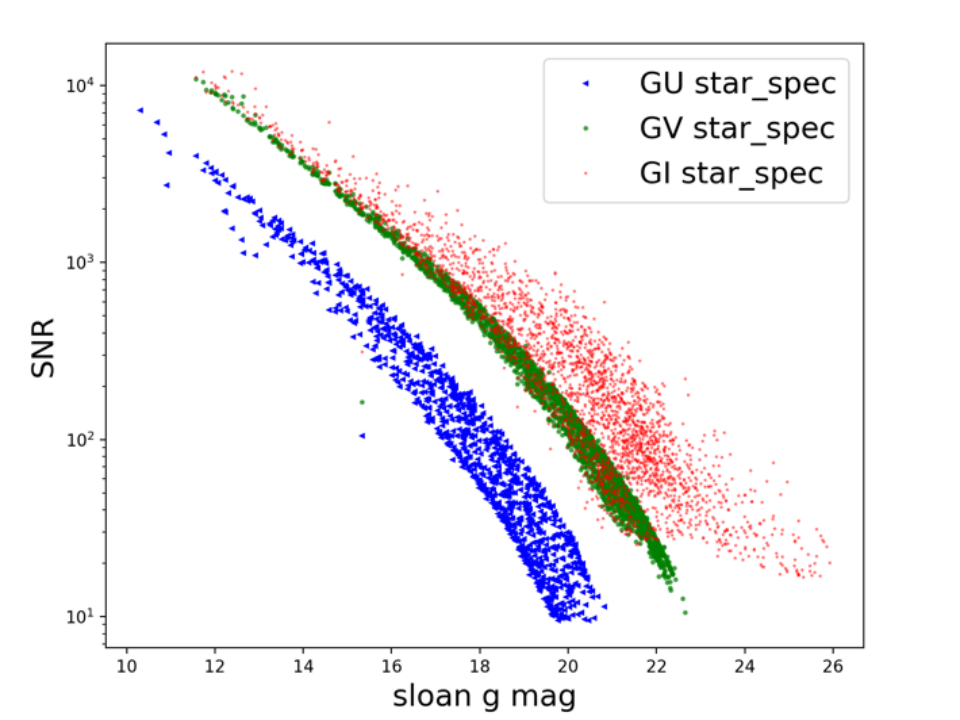}
        \caption{stellar}
        \label{fig_sim_snr2}
    \end{subfigure}
    \caption{The corresponding relationship between the brightness and signal-to-noise ratio of the target source of the detected 1st order spectrum.}
    \label{fig_sim_snr}
\end{figure*}

Fig.~\ref{fig_sim_snr} illustrates the relationship between the brightness and signal-to-noise ratio (SNR) of the target source for the 1st order spectrum. As evident from the figure, the SNR of the detected spectrum follows the order GI > GV > GU. Notably, the SNR of GU is significantly lower compared to the other two bands. Consequently, not only is the number of GU bands limited, but the measurement accuracy is also reduced due to its lower SNR.

\section{Summary}
\label{sect:summary}

In this study, the simulation of the CSST slitless spectrum primarily focuses on modeling the spectral dispersion characteristics of the true spectrum. These characteristics are influenced by multiple factors, including the groove density of the grating, the Point Spread Function (PSF) of the optical system, the relative position between the grating and the focal plane, and the manufacturing process of the grating. Due to these factors and the large field of view (FoV) of the CSST, the image quality cannot be guaranteed to be uniform across all positions. Additionally, the large size of the grating makes it challenging to ensure consistent spectral dispersion characteristics at every position. Therefore, it is necessary to sample multiple positions on the same grating to accurately fit the spectral dispersion characteristics.

Currently, there are no measured spectral data available for the CSST. As a result, this study adopts an optical simulation approach, utilizing optical software to generate spectral data at different positions on the grating based on the design parameters of the optical system and the grating. In future laboratory tests, a uniformly perforated coronagraphic mask will be placed over the grating of the detector to be tested. This setup will allow light to pass through the mask, creating uniformly distributed point sources across the entire grating area. By illuminating the grating with narrowband light sources of different wavelengths, the dispersed positions of various wavelengths can be captured on the detector. Using the fitting method described in this paper, a configuration file characterizing the grating's spectral dispersion properties can be generated. By fitting the measured data, the spectral characteristics of the CSST slitless spectrometer will be more accurately and realistically represented.

\normalem
\begin{acknowledgements}
This work is supported by the project of the CSST scientific data processing and analysis system of the China Manned Space Project. XZ acknowledges the support of National Natural Science Foundation of China (grant No.U2031143).

\end{acknowledgements}
  
\bibliographystyle{raa}
\bibliography{bibRAA-2025-0055.R1}

\end{document}